# Estimating Waning of Vaccine Effectiveness: a Simulation Study

**Author Affiliations:**

*Ariel Nikas*, Department of Microbiology and Immunology, Emory University School of Medicine, Atlanta, United States of America

*Hasan Ahmed*, Department of Biology, Emory University, Atlanta, United States of America

*Veronika I. Zarnitsyna*, Department of Microbiology and Immunology, Emory University School of Medicine, Atlanta, United States of America

**Contact Information:** vizarni@emory.edu

## Abstract:

**Background.** Developing accurate and reliable methods to estimate vaccine protection is a key goal in immunology and public health. While several statistical methods have been proposed, their potential inaccuracy in capturing fast intra-seasonal waning of vaccine-induced protection needs to be rigorously investigated.

**Methods.** To compare statistical methods for vaccine effectiveness (VE) estimation, we generated simulated data using a multiscale agent-based model of an epidemic with an acute viral infection and differing extents of VE waning. We extended the previously proposed framework for VE measures based on the observational data richness to assess changes of vaccine-induced protection with time.

**Results.** While VE measures based on hard-to-collect information (e.g. exact timing of exposures) were accurate, usually VE studies rely on time-to-infection data and the Cox proportional hazard model. We found that its extension utilizing scaled Schoenfeld residuals, previously proposed for capturing VE waning, was unreliable in capturing both the degree of waning and its functional form and identified the mathematical factors contributing to this unreliability. We showed that partitioning time and including a time-vaccine interaction term in the Cox model significantly improved estimation of VE waning, even in the case of dramatic, rapid waning. We also proposed how to optimize the partitioning scheme.

**Conclusions.** Using simulated data, we compared different measures of VE for capturing the intra-seasonal waning of vaccine-induced protection. We propose an extension of the Cox model based on including a time-vaccine interaction term with further optimization of partitioning time. These findings may guide future analysis of VE waning in observational data.

## Background

Accurate estimation of the extent of waning of vaccine-induced protection over time is an important public health need. Epidemiological data shows that protection from the influenza vaccine and COVID-19 vaccines could wane intra-seasonally [1-4]. While vaccine effectiveness (VE) measured a month after the second dose for mRNA vaccines against SARS-CoV-2 was reported to be over 90% [5-7], the recent data have shown a resurgence of SARS-CoV-2 infection in vaccinated people approximately 6 months after vaccination [3], indicating relatively fast waning of vaccine-induced protection and raising a question about the necessity for single or potentially multiple booster shots. Similarly, the data have shown that protection provided by influenza vaccination may wane intra-seasonally [1, 2, 4, 8-10] with the odds ratio of being tested positive for influenza infection in one study increasing linearly by approximately 16% for each additional 28 days since vaccination [4]. We should note that for both influenza and COVID-19 vaccines, the decline in vaccine effectiveness is associated with declining levels of antibody and antigenic evolution away from the vaccine strain. While antibody decline is a general phenomenon, for many other pathogens, such as diphtheria and tetanus, antibody titers may be far above the threshold of protection for decades after immunization [11]. In contrast in the case of coronaviruses, natural infection, which usually gives a stronger and longer lasting protection in comparison to vaccines [12, 13], still allows individuals to be reinfected even with the same strain after 12 months [12, 14] indicating inherently short-lived protection to this type of pathogen.

In this study, we use multiscale, agent-based models to compare different methods for estimating the waning of VE. Although multiple methods for estimating waning of VE have been proposed, relatively few simulation studies have compared the accuracy of these methods. Specifically for methods using extensions of the Cox proportional hazards model for VE estimation, comparison studies are sparse [15, 16] although this model is used frequently [10, 17- 21]. We find that a commonly used method using scaled Schoenfeld residuals [22] is reasonably accurate in detecting the presence or absence of waning but may be unreliable in estimating the degree of waning. This unreliability arises because the method actually estimates an approximation of an approximation of the time-varying hazard ratio with both approximations potentially introducing substantial error. In contrast, we show that a relatively straightforward method, creating an optimized time-vaccine interaction, performed much better and can be easily implemented.

## Methods

We consider measures for vaccine effectiveness based off the established framework and terminology found in [23, 24]. Here we focus on the direct effect of the vaccine on susceptibility to infection ($VE_S$), excluding indirect effects such as herd immunity. In this framework, estimators are grouped into three levels based on degree of information needed.

Level 1 measures of vaccine effectiveness require the most detailed information. Level 1 relies on knowing the number of infections and exposures in both the vaccinated and unvaccinated groups. Except in controlled challenge experiments, this data is typically difficult to obtain exactly but may sometimes be extrapolated from known household exposures [25, 26].

Easier to obtain, level 2 measures of vaccine effectiveness use infection data from both groups along with the person-time at risk for each group. Instead of knowing exact exposures, this assumes that contact rates with infectious individuals per unit of time are approximately equivalent in both groups.

Level 3 measures utilize proportional hazards, with the Cox proportional hazards model named specifically by the established framework [23, 24]. The difference between level 2 and level 3 is relatively subtle. Level 3 only needs the order in which infections occurred as well as the vaccination status (and any other covariate information) for both the infected individual and the at-risk study population at the infection times. A comparison of these three levels is shown in Table 1. In practice, level 3 estimates are frequently used as the Cox proportional hazards model is conveniently implemented in statistics software and can easily handle censoring and multiple, even time-varying, covariates. In addition, a convenient test for waning utilizing the Schoenfeld residuals is available for this class of models.

In order to observe waning, we consider time-varying measures at all three levels. Time-varying level 1 and 2 estimates can be calculated over specified time periods or with moving averages to create a smoother appearance. There exists a standard method to extend the level 3 Cox model to be time-varying for vaccine effectiveness studies utilizing the scaled Schoenfeld residuals [10, 18]. Taking a local average, at a given time, of the scaled Schoenfeld residuals gives an estimate of the log of the hazard ratio (in this case comparing vaccinated to unvaccinated individuals) at that time and, hence, also an estimate of VE at that time.

While this local average is typically calculated using the LOESS algorithm or natural splines, both of which give estimates that are continuous over time, for analytical tractability and to aid in direct comparison between all levels we derive local averages by creating bins (time categories) with a minimum of 100 events each, example in Figure S1. We note that binning, local regression, and splines are all nonparametric methods which should converge to each other so long as model complexity is increased appropriately as sample size increases. As is shown in Figure S2, our method of smoothing is not substantially different from LOESS smoothing in this context (see Supplemental Materials for details). We use these same bins to derive estimates for all three levels. In addition to the level 3 method already described, we also consider another level 3 method in which an interaction between vaccination and the time categories is used as the independent variable of the Cox regression.

## Results

To compare how the levels perform under different potential vaccine study circumstances, we modeled four separate epidemic scenarios, each with 100,000 individuals with 40% vaccine coverage per [27]. We consider a 'leaky' vaccine that gives constant protection with VE of 80%, meaning per exposure a vaccinated individual has 20% of the chance of being infected compared to an unvaccinated individual, in a study where either all vaccinated individuals receive their vaccine on the same day or where vaccination occurs spread over a period of 30 days. We contrast this with a hypothetical leaky vaccine with waning protection decreasing from 100% to 0% protection over 60 days where, again, vaccination occurs either on a single day or spread over 30 days. We intentionally model dramatic waning in order to test the limits of the methods. This set up is shown in Figure 1, giving both the VE value over time for a vaccinated individual and the average VE over time for not-yet-infected vaccinees.

For each scenario, we ran 100 simulations where contacts are randomly generated, probability of infection is based on vaccination status, and reinfection cannot occur in the short intra-seasonal window. Due to the stochastic nature of our simulations, a representative simulation was chosen for each scenario based on average infection numbers and epidemic peak timing. All simulations used previously estimated parameters for influenza and were created in Julia version 1.3.1 [28]. All analysis was completed using R version 3.6.1 [29]. Further modeling details can be found in the Supplemental Material.

**Estimating Vaccine Effectiveness for Constant and Waning Protection**

For constant protection at 80% reduced chance of infection throughout the season, all levels behave reasonably accurately as seen in Figure 2 A-B. The spread of vaccination understandably affects the dynamics of infection but shows little effect on the estimation. Visually, the effect of the vaccine appears to be constant for all levels except for some outliers when infection rates are low.

However, as vaccine protection wanes, differences between the levels emerge and become more dramatic as vaccination is spread, as seen in Figure 2 C-D. Levels 1 and 2 both capture early season behavior and remain similar until infection numbers become low. When considering a single day of vaccination, the level 3 Schoenfeld residuals (SR) method underestimates early season behavior and generally behaves less accurately than either level 1 or 2 estimates, with all three behaving more erratically as infection numbers decrease. When vaccination is spread over time, the SR method loses all accuracy except for at the very peak of infection.

We especially focus on the scenario where vaccination is spread, as many real world studies are based on observation or rolling enrollments. While VE studies often utilize the SR method [10, 19, 30], other fields using the Cox model for survival analysis have considered including a time-covariate interaction instead [31, 32]. In addition to the level 3 method already described, we also consider another level 3 method in which an interaction between vaccination and the time categories is used as the independent variable of the Cox regression. This time-vaccine interaction method (TVI), as seen in Figure 3, can improve accuracy in the VE estimate. In the next section, we explore the mathematical differences between the two methods, emphasizing the approximations that can lead to inaccurate results in the SR method.

**Mathematical Factors Contributing to Inaccuracy of SR Method**

In our simulations, when vaccination is spread over time and waning, the level 3 TVI method was approximately as accurate as the level 1 and 2 estimates whereas the level 3 SR method showed very large errors. Here we investigate the source of this error.

Given a study period $j$ (time bin $j$), we consider the following equation,

$$\frac{1}{n_j}\sum_{i=1}^{n_j} V_{j,i} = \frac{1}{n_j}\sum_{i=1}^{n_j} \frac{e^{\hat{\beta_{TVI,j}}Z_{j,i}}}{e^{\hat{\beta_{TVI,j}}Z_{j,i}}+(1-Z_{j,i})}, \tag{1}$$

where $n_j$ is the number of events in the time period $j$, $V_{j,i}$ is an indicator variable for whether the $i$-th infection (in time period $j$) is in a vaccinated individual, $Z_{j,i}$ is the proportion of the never infected population that is vaccinated at the time of that $i$-th infection, and $\hat{\beta}_j$ is the coefficient, estimating the natural log of the hazard ratio, to be calculated. Here, the left-hand side of the equation is the observed fraction of infections that occurred in vaccinated individuals during the time period $j$, and the right-hand side is the fraction expected by the Cox model. Solving this equation exactly corresponds to implementing the TVI method. We note that this value of $\hat{\beta}_j$ also corresponds to maximizing the partial likelihood function for the Cox model. For ease of legibility, we drop the $1/n_j$ from both sides of the equation in the following steps.

Taking a first-order Taylor series expansion centered at $\hat{\beta}$ of the right-hand side of the equation yields,

$$\sum_{i=1}^{n_j} V_{j,i} = \sum_{i=1}^{n_j} \frac{e^{\hat{\beta} Z_{j,i}}}{e^{\hat{\beta} Z_{j,i}} + (1 - Z_{j,i})} + \frac{e^{\hat{\beta} Z_{j,i}} (1 - Z_{j,i})}{\left(e^{\hat{\beta} Z_{j,i}} + (1 - Z_{j,i})\right)^2} \left(\widehat{\beta_{TS,J}} - \hat{\beta}\right). \tag{2}$$

Here $\hat{\beta}$ is the coefficient estimated by the simple Cox model without a time-vaccine interaction. If we introduce a weighting term,

$$W_{j,i} = \frac{\left(e^{\hat{\beta} Z_{j,i}} + (1 - Z_{j,i})\right)^2}{e^{\hat{\beta} Z_{j,i}} (1 - Z_{j,i})}, \tag{3}$$

the previous equation can be modified as shown,

$$\sum_{i=1}^{n_j} W_{j,i} V_{j,i} = \sum_{i=1}^{n_j} W_{j,i} \left( \frac{e^{\hat{\beta} Z_{j,i}}}{e^{\hat{\beta} Z_{j,i}} + (1 - Z_{j,i})} + \frac{e^{\hat{\beta} Z_{j,i}} (1 - Z_{j,i})}{\left(e^{\hat{\beta} Z_{j,i}} + (1 - Z_{j,i})\right)^2} \left(\widehat{\beta_{SRTV,J}} - \hat{\beta}\right) \right)$$

$$= \sum_{i=1}^{n_j} W_{j,i} \left( \frac{e^{\hat{\beta} Z_{j,i}}}{e^{\hat{\beta} Z_{j,i}} + (1 - Z_{j,i})} \right) + \left(\widehat{\beta_{SRTV,J}} - \hat{\beta}\right) \tag{4}$$

The above corresponds to the scaled Schoenfeld residuals method if the residuals are calculated as first derived in [22] (henceforth, we called this the SRTV method). We note that the above weighting is algebraically necessary to easily extract residuals but further removes this equation from Equation 1. The reciprocal of this weighting term is often referred to as the 'variance', as it roughly corresponds to the variance of $V_{j,i}$ and, in this case, it changes over time. Commonly, this time-varying

'variance' is replaced by a constant average 'variance' $\widehat{\text{Var}}(\hat{\beta})n$ in the manner shown below.

$$\sum_{i=1}^{n_j} V_{j,i} = \sum_{i=1}^{n_j} \frac{e^{\hat{\beta} Z_{j,i}}}{e^{\hat{\beta} Z_{j,i}} + (1 - Z_{j,i})} + \frac{\widehat{\beta_{SR,J}} - \hat{\beta}}{\widehat{\text{Var}}(\hat{\beta})n} \tag{5}$$

where *n* is the total number of events observed across all time categories. This equation corresponds to the SR method using the scaled Schoenfeld residuals as implemented in R's `survival` package. We note that Equation 5 can be derived directly from Equation 2. So, any imprecision introduced by the weighting in Equation 4 is removed but at the cost of another assumption. Hence, both Equations 4 and 5 can be viewed as approximations of Equation 2 which is itself an approximation of Equation 1.

We find that simply taking the Taylor series expansion (Equation 2, TS method) introduces substantial error, shown in Figure 4. In this situation, the weighting introduced in Equation 4 has practically no effect. However, we cannot rule out that in other studies, such as ones with much smaller sample size, this weighting may be influential. Notably, by far the largest source of error was introduced by replacing the time-varying 'variance' with a constant (Equation 5).

**Accuracy of Waning Detection Using the SR Test**

As previously stated, level 3 has a statistical test for constant versus time-varying which uses the correlation between the Schoenfeld residuals and time. When applied to simulations without waning, this test erroneously detected waning in 7% of cases for simulations with 1 day spread and 6% of cases for simulations with 30 day spread. This is near the expected false positive rate of 5%, and overall the *p-value* distribution, as shown in Figure S3, falls near the expected uniform distribution. The test correctly detects waning in 100% of the simulations that wane, which is not surprising given the large sample size and intentionally dramatic waning. Hence, this test performs appropriately in our simulations even as the degree of waning is sometimes very poorly estimated by the scaled version of the Schoenfeld residuals.

**Optimization of VE Estimation**

While the level 3 TVI method was reasonably accurate, we used an arbitrary, though not unreasonable, number of events to partition time for the calculation of each method.

To find optimized partitions for the TVI method, we considered various combinations of

minimum number of days per bin and minimum number of events per bin to create

partitions for several R0 values as shown in Figure 5. Specifics on bin creation can be

found in the Supplemental Materials. All simulations have the same waning of 100-0%

with vaccination spread over 30 days.

Because we know the true value of protection, we can calculate the root mean square

error (RMSE) for each combination; however, an analyst using a real world data set

would not have such knowledge. We found that the combination with the minimum

Akaike information criterion (AIC), which requires no prior knowledge, generally corresponds to low RMSE and therefore can be used to create an optimal or near optimal estimate. The difference between our arbitrarily chosen 100 event minimum binned TVI model and the optimal AIC binned model is shown in Figure 6. This optimized level 3 TVI estimate corrects the underestimates of early season behavior that the other level 3 methods display, closely follows the functional form of vaccine protection even as infection numbers diminish and is additionally easy to compute.

**Linear Interpolation versus Step Function**

For figures showing VE estimates, we used simple linear interpolations to connect the bins, but the underlying models are actually piecewise constant step functions, a biologically implausible pattern that may confuse certain readers. Hence, we considered the effect of replacing the underlying step function of the optimized time-vaccine interaction method with linear interpolation using the mean event time and VE estimate for each bin and then connecting those points. Linear interpolation performs both qualitatively and quantitatively better than the step function in this circumstance, improving RMSE from 2.85% to 1.55% when using the optimized bins (Figure S4). As such, we recommend linear interpolation to gain continuity and additional accuracy.

## Discussion

Correctly estimating vaccine effectiveness and the extent to which it wanes is a key goal in immunology and public health, and several different methods for capturing the extent and form of waning have been proposed. We compared methods from the framework proposed in [23, 24] using a simulated seasonal epidemic of an acute viral infection. The level 1 and level 2 methods considered performed reasonably well. However, we found that the level 3 scaled Schoenfeld residuals (SR) method, commonly used and recommended for vaccine effectiveness studies, while potentially quite accurate in some circumstances, has difficulty capturing waning in some other scenarios. Moreover, the statistical justification for the SR method is questionable since the local average of the scaled Schoenfeld residuals, in general is not and does not asymptotically approach the maximum partial likelihood value (Equation 1 vs Equation 5). In stark contrast, in our simulations the Schoenfeld residuals test performed appropriately with regards to rejecting the null hypothesis of no waning. This discrepancy possibly arises because the null hypothesis of no waning is a special case in which the previously noted inconsistency, between the local average of the scaled Schoenfeld residuals and the maximum partial likelihood value, (asymptotically) vanishes. We show that a straightforward approach, creating time categories and adding a time-vaccine interaction term as a predictor variable, the TVI method, performs accurately in all scenarios considered and sometimes much better than the SR method. Optimizing the time categories and adding linear interpolation further increases accuracy, yielding estimates that closely follow the functional form of vaccine protection in the population. This method, as it uses only standard statistical techniques, should be easy to implement in any statistical programming language that includes Cox regression. The most computationally difficult step, optimizing the time categories, can potentially be avoided or simplified by using prior knowledge to fix minimum number of days or minimum number of events or by optimizing only on minimum number of events per bin.

Vaccine effectiveness papers have continued to recommend scaled Schoenfeld residuals as a method to estimate VE over time [10, 17-21]. For example, a recent paper comparing statistical methods [16], concludes this SR method is the best of the ones they tested. On this note, it makes sense to consider why our conclusions differ from those of previous papers. First, since the approximations inherent in the SR and SRTV methods are sometimes very accurate and at other times results in larger error, it seems likely that other papers only considered scenarios where the approximations are accurate. Secondly, and perhaps more importantly, some of the other papers did not consider the full potential of the TVI method. For example [16] only considers a highly parametric interaction, vaccination by log(time).

Finally, we examine some of the limitations of our study. We do not consider the thorny issues of heterogeneity in exposures, heterogeneity in vaccine effectiveness, and unmeasured confounders. Because this paper is proof of concept, we do not consider age stratification or other heterogeneities. We deliberately used a large sample size in our simulations and smaller samples sizes would have a different bias variance tradeoff. However, reducing the sample size to 1000 infections gave similar results. Additionally, we intentionally focused on relatively standard methods. We did not consider innovative and newer methods such as [33] or [34]. It is likely that some of these methods may have superior performance especially at smaller sample sizes. Instead we provide a baseline for what can be achieved using established statistical methodology while avoiding the unreliability of the SR method.

***Financial support.*** This work was supported by the National Institutes of Health (grant number U01HL139483).

## Tables:

Table 1: Vaccine effectiveness measures for susceptibility, where V represents the vaccinated group and U the unvaccinated.

| Measure | Formula | Required Data |
|---------|---------|---------------|
| Level 1 | $VE = 1 - \dfrac{Infections_V / Exposures_V}{Infections_U / Exposures_U}$ | Time of Exposures, Vaccinations, & Infections |
| Level 2 | $VE = 1 - \dfrac{Infections_V / Person\text{-}time_V}{Infections_U / Person\text{-}time_U}$ | Time of Vaccination and Infection |
| Level 3 | $VE = 1 - Hazard\ Ratio$ | Order of Vaccination and Infection |

## Figures:

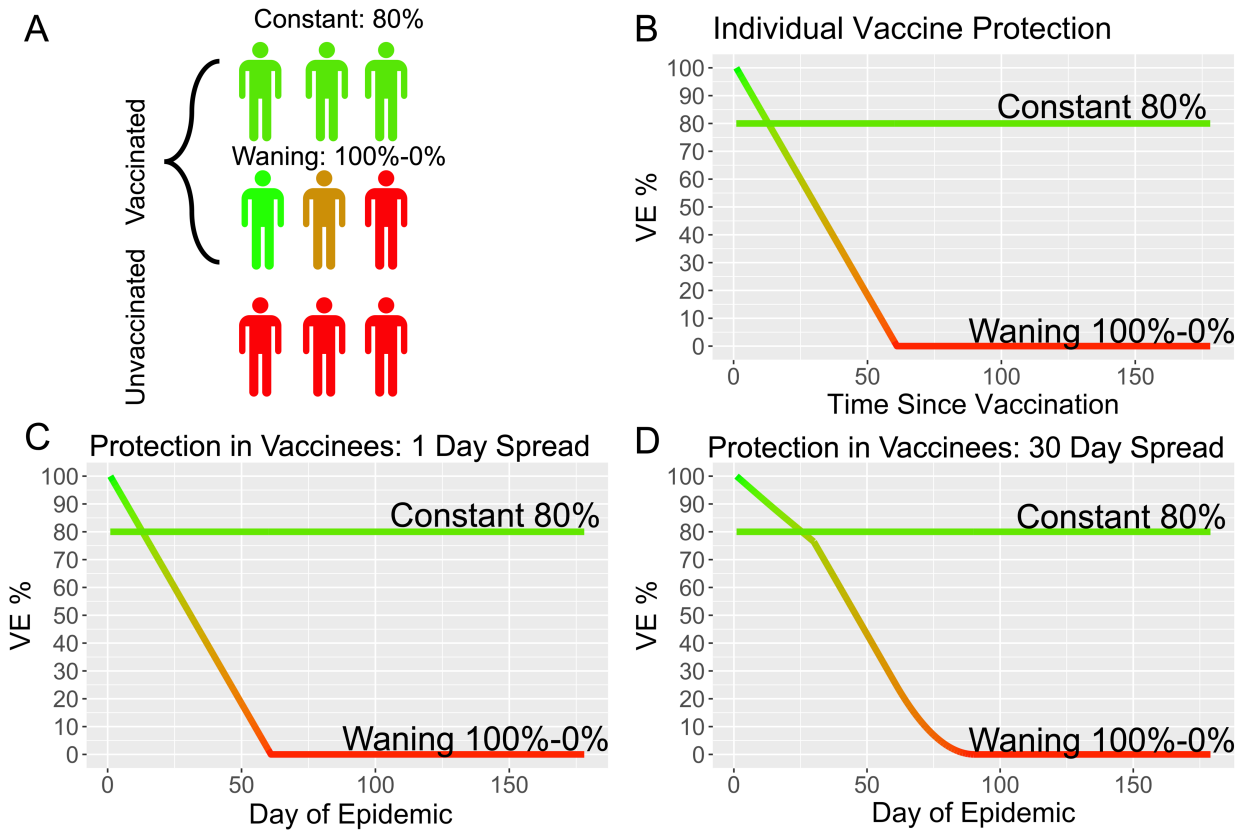

Figure 1: **Comparing the true values at the individual and population level for four scenarios.** As shown in Panel A, we consider four scenarios where either the vaccine protection remains constant at 80% or where protection wanes from 100% to 0% over 60 days with and without spread of vaccination. The individual level of protection is shown in Panel B. This would also be the population average if vaccination occurs on a single day, as shown in Panel C. When vaccination is spread over 30 days, individuals will have differing extents of protection, and, therefore, the average protection in the susceptible vaccinated population has the form as shown in Panel D. See formula in Supplemental Material (Equation SM 1).

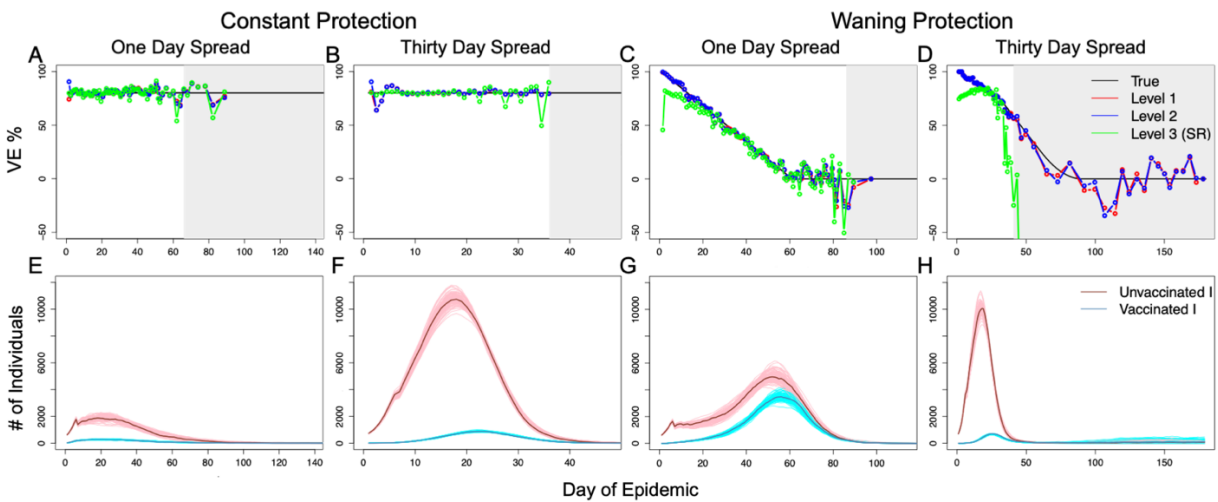

Figure 2: **Estimating vaccine effectiveness for constant and waning protection.** When vaccine protection is constant at 80% for one day and thirty day spread of vaccination (Panels A and B, respectively), all levels are reasonably accurate when infection numbers are high as can be seen in their corresponding epidemic dynamics in Panels E and F. If vaccination is spread, early results should be considered with caution due to the extremely small relative size of the vaccinated group which can lead to lack

of infections and exposures in that group. However, when vaccine protection is waning (Panels C and D), complications arise especially for level 3, which underestimates early season behavior. This is especially clear when vaccination is spread, as in Panel D, with the level 3 Schoenfeld residual (SR) method only estimating accurately at the very peak of infection. In Panels A-D, light grey boxes display the region where total daily infections are low (<60 as suggested in [22]), and x-axis ends on day of last infection.

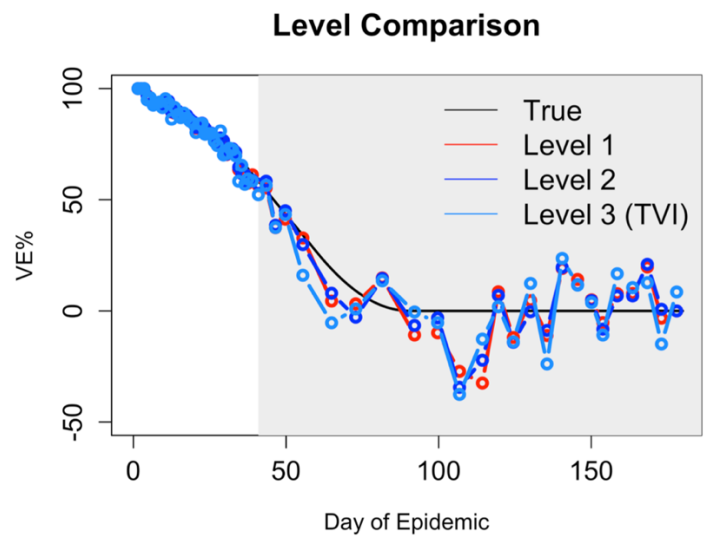

Figure 3: **Inaccuracy not due to insufficient information for level 3**. While the SR method for calculating level 3 values may fail in certain circumstances, level 3 estimates are not inherently inaccurate. Using a time-vaccine interaction (TVI) method, as defined in § *Mathematical Factors Contributing to Inaccuracy of SR Method*, estimates for level 3 are quite similar to level 1 and 2 estimates.

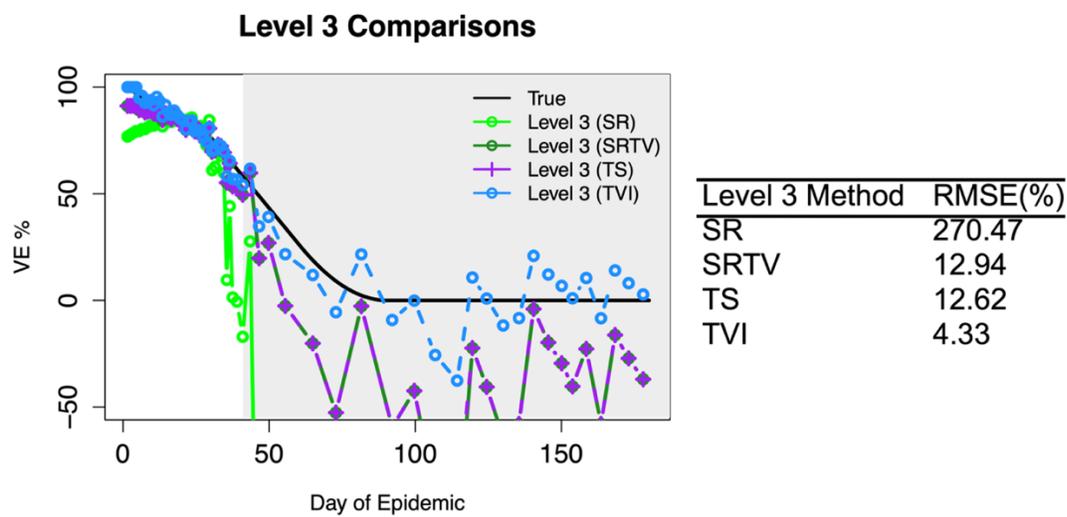

Figure 4: **Approximations of level 3 methods can lead to substantial error**. With each approximation from the time-vaccine interaction (TVI) method, error is compounded. With the first approximation coming from the Taylor series (TS) and its weighted version, the SR method with time-dependent variance (SRTV), error increases especially when event number is low. This error increases further under the additional assumption that variance is fixed, giving the standard Schoenfeld residual (SR) method.

This error is obvious qualitatively and also when quantified by the root mean square error (RMSE).

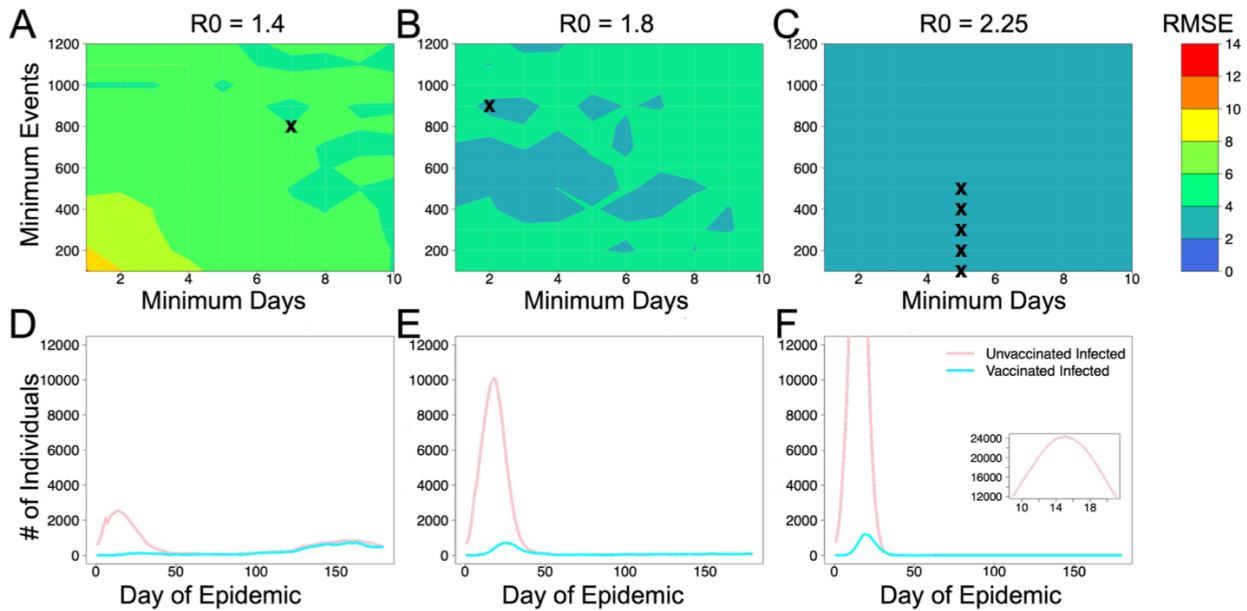

Figure 5: **Rather than using simple 100 event bins, the time vaccine interaction method can be further improved.** Because we know the expected value for each simulation VE(t), we can calculate the root mean square error (RMSE) for the time-stratified model against the expected value but in a real-world study this would not be the case. However, over a variety of R0 values, we find the minimum AIC, shown as **X**'s on Panels A-C, corresponds well to where low root mean square error is found. So, despite the very different dynamics in each of these systems, a simple check over multiple minimums for events and days for minimum AIC should yield a good result.

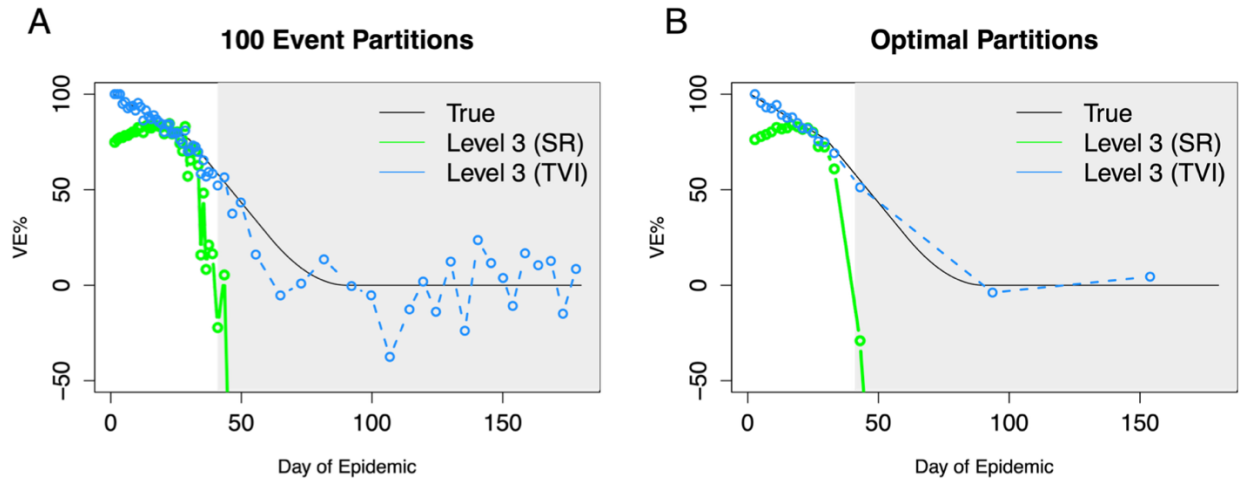

Figure 6: **Optimizing the time categories gives an excellent visualization of the true form**. In Panel A, simple 100 event partitions are used while in Panel B partitions are a minimum of 6 days with 700 events per the minimum AIC for this simulation. This offers an improvement to the estimate even in such an extreme case using a relatively simple method. Optimizing the partitioning lowered RMSE from 4.33% to 2.85%.